\documentclass[reprint,superscriptaddress,prl]{revtex4-1}
\usepackage{natbib}
\usepackage{color}
\usepackage{amsfonts,amssymb,amsmath}
\usepackage{graphics,graphicx,lineno,subfig,caption}
\usepackage{amsmath, amsthm, amssymb}
\begin{document}
\title{The influence of the choice of post-processing method on Bell inequalities}
\author{Miko\l aj Czechlewski}
\email{mczechlewski@inf.ug.edu.pl}
\affiliation{Institute of Informatics, National Quantum Information Centre, Faculty of Mathematics, Physics and Informatics, University of Gda\'nsk, Wita Stwosza 57, 80-308 Gda\'nsk, Poland}
\author{Marcin Paw\l owski}
\affiliation{Institute of Theoretical Physics and Astrophysics, National Quantum Information Centre, Faculty of Mathematics, Physics and Informatics, University of Gda\'nsk, Wita Stwosza 57, 80-308 Gda\'nsk, Poland}
\date{\today}
\begin{abstract}
In every experimental test of a Bell inequality, we are faced with the problem of inefficient detectors. How we treat the events when no particle was detected has a big influence on the properties of the inequality. In this work, we study this influence. We show that the choice of post-processing can change the critical detection efficiency, the equivalence between different inequalities or the applicability of the non-signaling principle. We also consider the problem of choosing the optimal post-processing strategy. We show that this is a non-trivial problem and that different strategies are optimal for different ranges of detector efficiencies.
\end{abstract}

%\pacs{1.2}
%\keywords{entanglement; Bell inequalities;non-locality}

\maketitle
\section{Introduction}\label{s1}
Undoubtedly, Bell's theorem
is one of the most significant achievements in the foundations of quantum mechanics \cite{Bell1964,Einstein1935}. After its publication, scientists have introduced many more Bell-type inequalities \cite{Clauser1969,Clauser1974,Svetlichny1987,Braunstein1990,Eberhard1993,Collins2002,Collins2004} with different properties. Some of these properties measure their usefulness for various tasks like quantum key distribution or disproving the possibility of a Local Hidden Variable (LHV) description of the world.

The first experimental violation of one of them was demonstrated already in 1972 \cite{Freedman1972}. However, it did not exclude the possibility of LHV model. This is due to the fact that this experiment did not overcome two main obstacles arising in such tests \cite{Larsson2014}: (a) avoiding the exchange of information about the measurement settings between particles; and (b) detecting particles with a probability high enough. If the detection efficiency is too low or one particle can have access to the choice of the setting of the other one, it is possible to reproduce quantum correlations using a LHV model. These two problems are known in the literature as the locality \cite{Bell1987} and the detection loophole \cite{Pearle1970,Santos1992}, respectively.

The locality loophole was first closed by Aspect \cite{Aspect1982} but in his experiment, the one related to detection was not. Closing it became an important problem not only because of the foundational issues mentioned above but also because loophole-free violation of local realism is a necessary condition for device-independent quantum information processing \cite{Acin2007}. Recently, the first experiment to close both these loopholes at the same time was reported \cite{Hensen2015} and other groups followed soon afterwards \cite{Giustina2015,Shalm2015}. Interestingly, the authors of \cite{Hensen2015} and \cite{Giustina2015} report violating different inequalities: CHSH and CH, respectively. These two have the same number of involved parties, settings and outcomes. From a practical point of view, the main difference between them is their vulnerability to detection efficiency loophole.

In \cite{Clauser1969} it was shown that for CHSH inequality the lower bound for detector efficiency needed to close the loophole is $82.8\%$. For CH \cite{Clauser1974} Eberhard has shown it to be $66.7\%$ \cite{Eberhard1993}. It is very common in the literature to make a clear distinction between these two inequalities, eg. the authors of \cite{Clauser1969} strongly stress it. Yet, it is also known that they are equivalent \cite{Fine1983}! One of the aims of this paper is the explanation of this paradox. We show that the reason for it lies in post-processing of experimental data, which is a necessity if the detectors are not 100\% efficient. For some types of post-processing, the equivalence is preserved while for the others it is not and different critical detection efficiencies appear. In the rest of the paper, we study other effects that arise in inequalities more complex than CHSH.

\section{The strategies of post-selection}\label{s2}

In this paper we consider two-partite Bell inequalities with binary outcomes. We will call the parties Alice and Bob with inputs $x$ and $y$ respectively. Their outcomes are $a\in\{0,1\}$ and $b\in\{0,1\}$.

Because of imperfections of detectors in a real Bell type experiment, there are always rounds in which some detector (or detectors) does not produce any outcome, in other words, does not "click". Therefore, it is a relevant question to ask: How should one treat these rounds? One of the possible solutions to this problem is to assign a new value, different from 0 and 1 to the no clicking outcome. However, this way is not satisfactory because:
\begin{description}
\item[(a)] A new outcome in an experiment requires a new Bell's inequality, which the coefficients are not known in advance.
\item[(b)] We often need a particular form of the inequality when it is used in some information processing protocol \cite{Vazirani2014}. Adding a new outcome changes that form which can now be useless for our purposes.
\end{description}
The problems mentioned above cause that experimentalists choose one of the two strategies \cite{Brunner2007,Eberhard1993}:
\begin{itemize}
\item THE DISCARD STRATEGY\\
If in any round of an experiment the detector of at least one party does not click, they discard this round from their statistical data. This strategy causes that the classical value of a tested Bell inequality to increase and the quantum value does not change.
\item THE ASSIGNMENT STRATEGY\\
If in any round of an experiment, Alice's or Bob's detector does not click that party assigns some output value (0 or 1) in this round of the experiment. This strategy causes that the classical value of a tested Bell inequality does not change and the quantum value decreases.
\end{itemize}
It is crucial to note that the discard strategy may imply signaling in the estimated joint probability distribution. One can demonstrate this by the following example:
Let us imagine that one has devices described by the joint conditional probability distribution $P(ab|xy\lambda)$, where $x,y\in\{0,1\}$ and $\lambda\in\{0,1\}$ is a hidden variable:
\begin{eqnarray}
P(a,b=0|xy\lambda)&=&\delta_{a,\lambda}\ .\label{joinProbHV}
\end{eqnarray}
The joint observed probability distribution for Alice and Bob is:
\begin{eqnarray}
P(a,b=0|xy)&=&\sum\limits_{\lambda}p_{\lambda}\delta_{a,\lambda}\ ,\label{joinProb}
\end{eqnarray}
where $p_{\lambda}$ is probability distribution of the hidden variable. Probability distribution is called non-signalling if the marginal probabilities of one party do not depend
on the input of the other, i.e.:
\begin{eqnarray}
\forall_{y,y'}\quad P^{A}(a|x)&=&\sum\limits_{b=0}^{1}P(ab|xy)=\sum\limits_{b=0}^{1}P(ab|xy^{'})\label{e1}\\
\forall_{x,x'}\quad P^{B}(b|y)&=&\sum\limits_{a=0}^{1}P(ab|xy)=\sum\limits_{a=0}^{1}P(ab|x^{'}y)\label{e2}
\end{eqnarray}
One can easily check that (\ref{joinProb}) satisfies these conditions since the marginal probabilities read:
\begin{eqnarray}
P^{A}(a|x)&=&\sum\limits_{\lambda}p_{\lambda}\delta_{a,\lambda}=p_{a}\ , \label{marginals1}\\
P^{B}(b|y)&=&\delta_{b,0} \label{marginals2}\ .
\end{eqnarray}

Let us now modify the probability distribution by allowing the devices not to click in some instances. While it never happens for Alice, Bob's device clicks if and only if $y=\lambda$. We introduce a variable $c_B$ which is equal $1$ if Bob's device did click and 0 otherwise. Its probability distribution reads:
\begin{eqnarray}
P(c_{B}=1|xy\lambda)=\delta_{y,\lambda}\ .
\end{eqnarray}
The statistics obtained after discarding all the rounds with $c_B=0$ are:
\begin{eqnarray}
P(a,b=0|xyc_{B}=1)&=&\frac{P(a,b=0,c_{B}=1|xy)}{P(c_{B}=1|xy)}\ ,\label{discProb}
\end{eqnarray}
where:
\begin{eqnarray}
P(a,b=0,c_{B}=1|xy)&=&\sum\limits_{\lambda}p_{\lambda}\delta_{a,\lambda}\delta_{y,\lambda}= p_y \delta_{a,y},\label{discProbElements}\\
P(c_{B}=1|xy)&=&\sum\limits_{\lambda}p_{\lambda}\delta_{y\lambda}=p_y \nonumber .
\end{eqnarray}
Formula \ref{discProb} becomes then:
\begin{eqnarray}
P(a,b=0|xyc_{B}=1)&=&\frac{p_y \delta_{a,y}}{p_y}= \delta_{a,y}\label{discProbStrict} .
\end{eqnarray}
and Alice's marginal $P^{A}(a|x)= \delta_{a,y}$ probability starts to depend on $y$ which violates condition (\ref{e1}).

A convenient way of presenting probability distributions allowed by a theory is to assigning a point in a 16 dimensional space to each of them. Point's coordinates are the values of $P(ab|xy)$ and every combination of the values $a,b,x$ and $y$ represents a dimension. Bell inequalities, being linear combinations of $P(ab|xy)$ are then directions in this space. For local and non-signalling theories the sets of allowed probability distributions are polytopes, while the structure of the quantum region is much more complex. Figure \ref{fig1a} based on \cite{Branciard2011} shows how the discard strategy influences CHSH inequality, with local polytope extending outside the non-signalling one.

On the other hand, the assignment strategy respects the non-signaling constraints. It follows directly from the fact that assigning an outcome is a local operation while discarding the whole round when only one part did not click is not. The impact of the assignment strategy on the space of conditional probabilities is shown in figure \ref{fig1b} and \ref{fig1c}. Especially, plot \ref{fig1c} needs explanation. The classical region stretches beyond the quantum one because we assume that the detectors behave differently in different theories.  For the quantum strategy the detectors are honest, which means that they do not click because of their physical imperfections. On the other hand, the clicking of the detectors for the classical strategy can be governed by some local hidden variable.

This assumption gives an advantage to the classical strategy over the quantum one.

It is worth to emphasize that the adopted strategy of the post-selection in a Bell experiment with inefficient detectors is crucial because it has an influence on the value of the threshold detector efficiency.

\section{Equivalence between CHSH, CH and Eberhard inequalities}\label{s3}
For boxes having two inputs, two outputs and shared by two parties there are known three main types of inequalities for the testing of local realism: CHSH, CH and Eberhard (E). We want to stress that although for the discard strategy these three inequalities are different, for the assignment all are equivalent. The equivalence is shown in the next subsections.
\subsection{Equivalence between CHSH and CH inequalities}\label{ss31}
To show the equivalence between CHSH and CH inequalities let us put CHSH as a sum of probabilities:
\begin{eqnarray}
-2\leq\sum\limits_{x,y,a,b=0}^{1}(-1)^{xy}(-1)^{a\oplus b}P(ab|xy)&\leq&2\ .\label{e4}
\end{eqnarray}
Next we write the expression $\sum\limits_{a,b=0}^{1}(-1)^{a\oplus b}P(ab|xy)$ in the standard way:
\begin{eqnarray}
\sum\limits_{a,b=0}^{1}(-1)^{a\oplus b}P(ab|xy)&=&P(00|xy)-P(01|xy)\label{ee4}\\
&-&P(10|xy)+P(11|xy)\nonumber.
\end{eqnarray}
If we use the assignment strategy as a post-selection method we assume that in the Bell experiment a probability distribution has to obey non-signaling constrains \ref{e1} and \ref{e2}, which one can write in the following way:
\begin{eqnarray}
P(01|xy)=P^{A}(0|x)-P(00|xy)\ ,\label{eee4}\\
P(10|xy)=P^{B}(0|y)-P(00|xy)\ .\nonumber
\end{eqnarray}
When we put \ref{eee4} into \ref{ee4} and use
the normalisation condition $\forall x,y \sum\limits_{a=0}^{1}\sum\limits_{b=0}^{1}P(ab|xy)=1$ we can derive the relation:
\begin{eqnarray}
\sum\limits_{a,b=0}^{1}(-1)^{a\oplus b}P(ab|xy)&=&4P(00|xy)-\label{e5}\\-2P^{A}(0|x)&-&2P^{B}(0|y)+1\ .\nonumber
\end{eqnarray}
Substituting \ref{e5} to \ref{e4} one gets CH inequality:
\begin{eqnarray}
-1\leq P(00|00)+P(00|01)+P(00|10)-\label{e7}\\
-P(00|11)-P^{A}(0|0)-P^{B}(0|0)\leq 0\nonumber\ .
\end{eqnarray}
Note that to show this equivalence we used non-signalling constraints (\ref{eee4}). However, these do not have to hold for probability distributions obtained by discard strategy. This brings us to the first main result of the paper: The choice of postprocessing method is responsible for (in)equivalency of CHSH and CH inequalities.
\subsection{Equivalence between E and CH inequalities}\label{ss32}
In 1993 Eberhard \cite{Eberhard1993} derived an inequality which considered three possible outcomes of detectors: $0$, $1$ and $\emptyset$. The symbol $\emptyset$ stands for no clicking of a detector:
\begin{eqnarray}
P(00|00)&-&P(00|11)-P(01|01)-\label{e9}\\
-P(0\emptyset|01)&-&P(10|10)-P(\emptyset 0|10)\leq 0\ .\nonumber
\end{eqnarray}
One can treat the outcomes $1$ and $\emptyset$ as one outcome because both of them are opposite to $0$. Note, that this implicitly assumes that we are using the assignment strategy here. Hence, Alice and Bob's marginal probabilities are computed in the following way \cite{Khrennikov2014}:
\begin{eqnarray}
P^{A}(0|0)&=&P(00|01)+P(01|01)+\label{e10}\\
&+&P(0\emptyset|01)\nonumber\ ,\\
P^{B}(0|0)&=&P(00|10)+P(10|10)+\label{e11}\\
&+&P(\emptyset 0|10)\nonumber\ .
\end{eqnarray}
Next, if one transforms the relations \ref{e10} and \ref{e11} and puts them into \ref{e9} one gets \ref{e7}. Note that this equivalence also holds only for the assignment strategy.
\section{Post-selection strategies for $I_{3322}$ inequality}\label{s4}
\begin{figure*}[t]
\subfloat[]
{\includegraphics[scale=0.3]{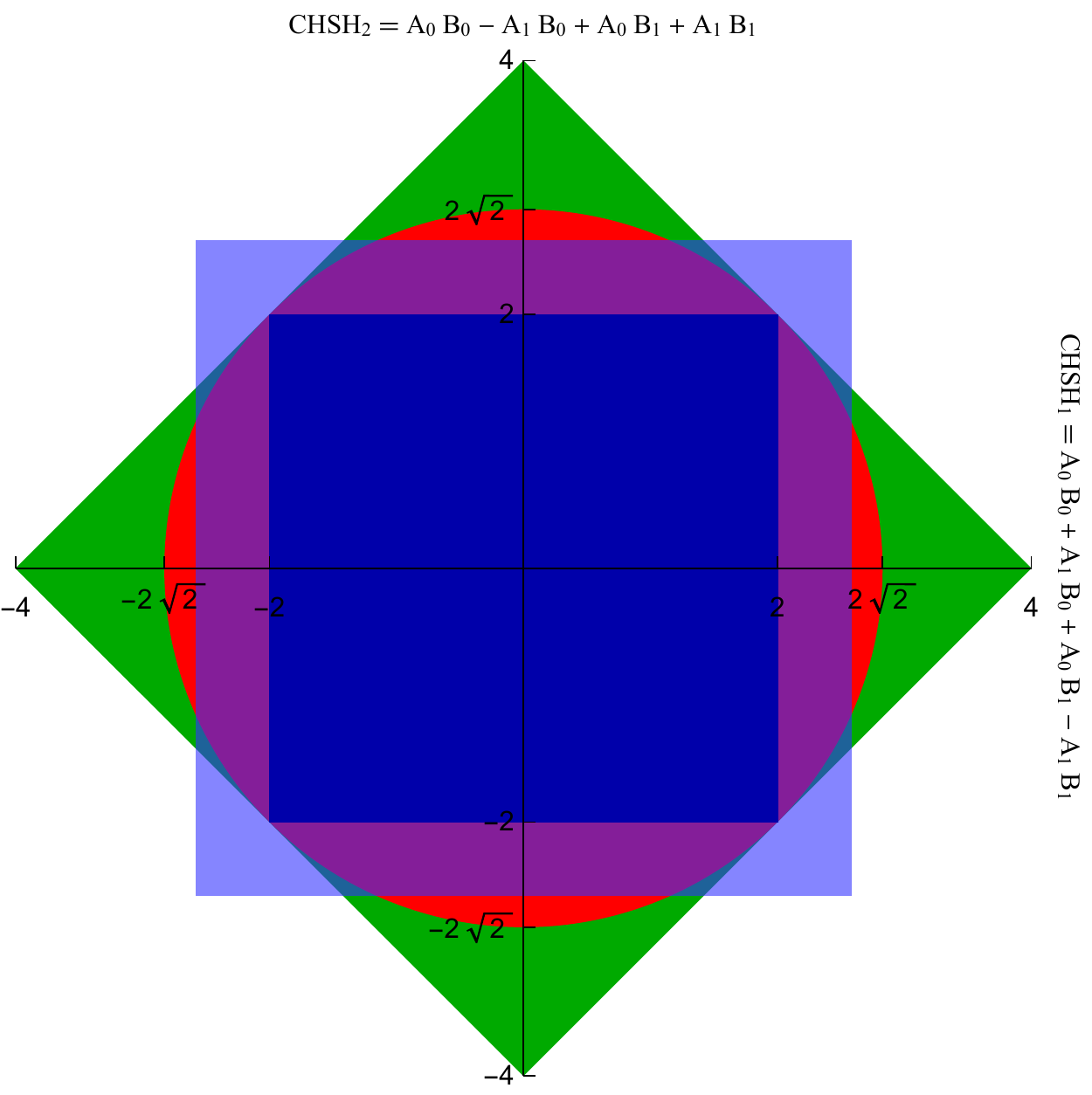}
\label{fig1a}}\qquad
\subfloat[]
{\includegraphics[scale=0.3]{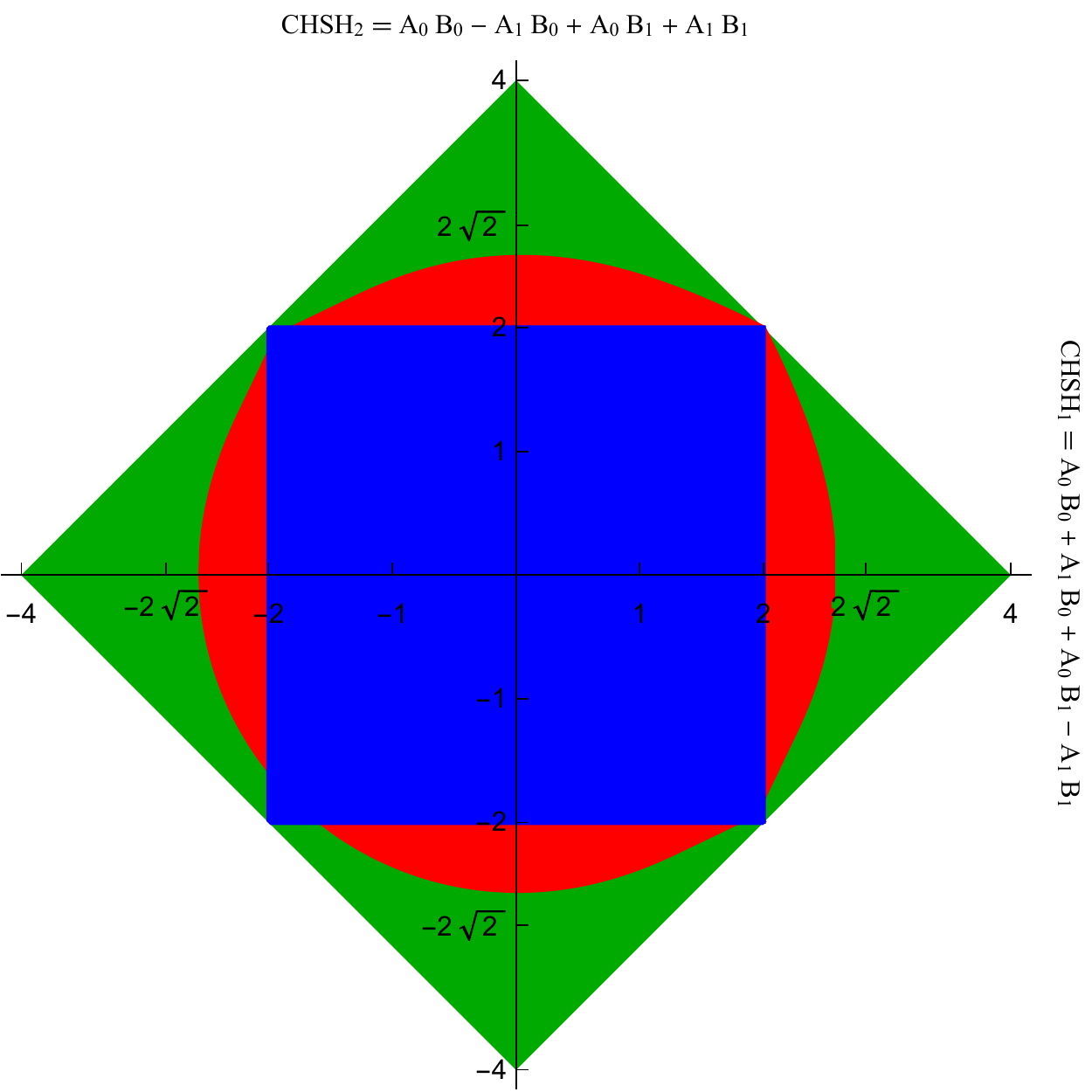}\label{fig1b}}\qquad
\subfloat[]
{\includegraphics[scale=0.3]{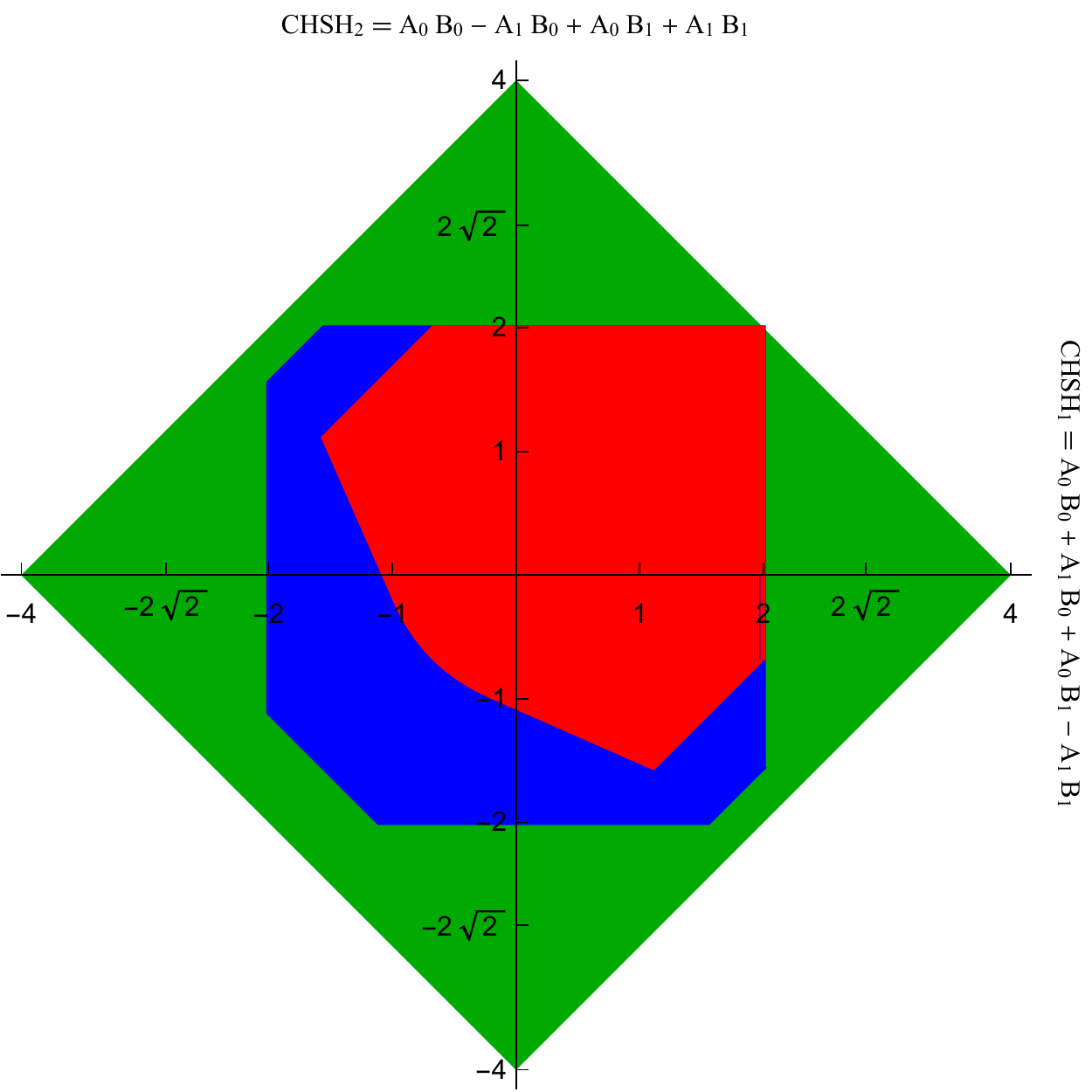}
\label{fig1c}}\\
\subfloat[]
{\includegraphics[scale=0.3]{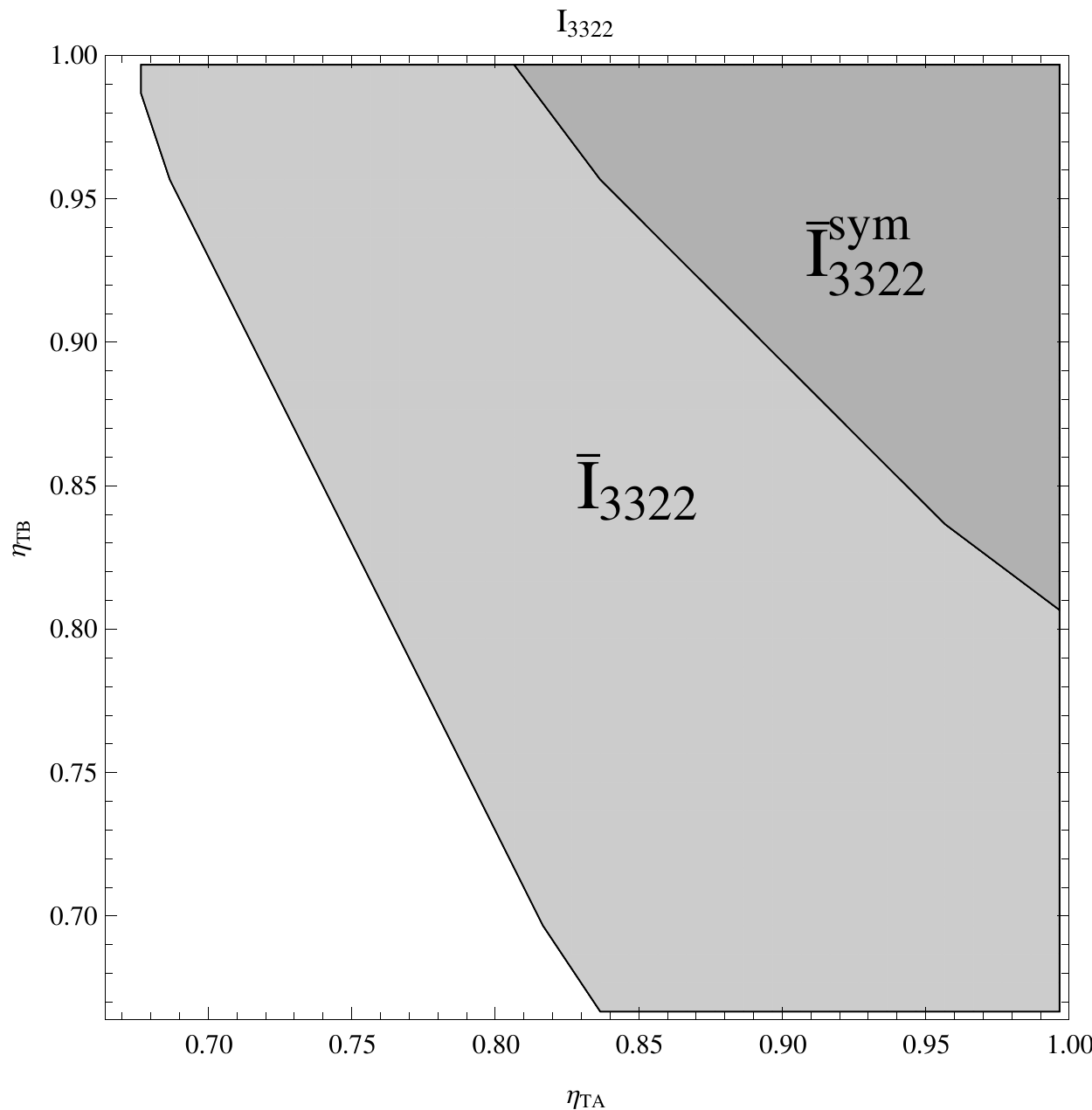}
\label{fig1d}}\qquad
\subfloat[]
{\includegraphics[scale=0.3]{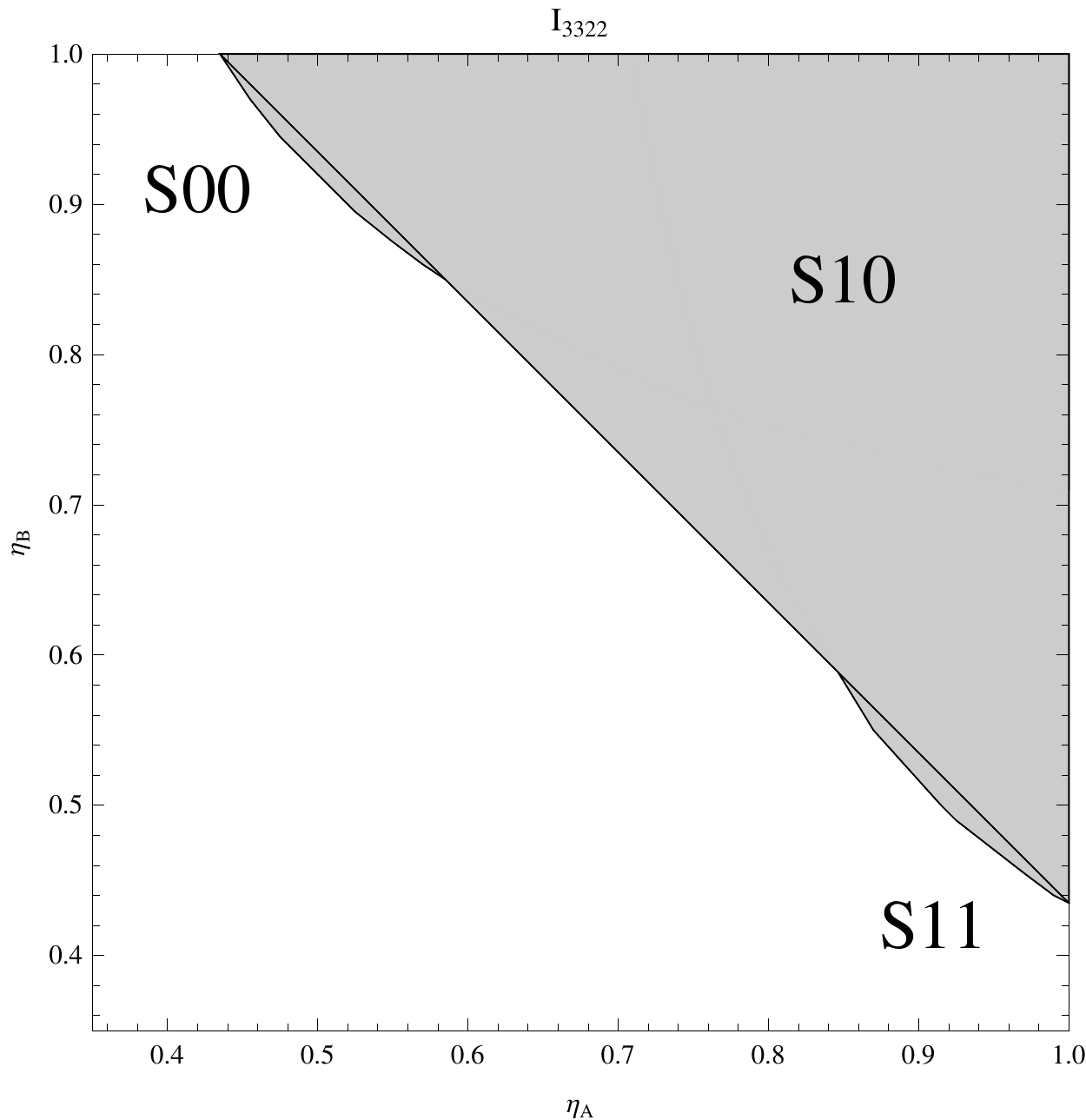}
\label{fig1e}}
\caption{
{\bf\ref{fig1a}}
The plot illustrating the correlations of CHSH inequality when the discard strategy is applied in the case of no clicking of the detectors for detector efficiency $\eta_{A}=\eta_{B}=\eta=0.88$ (light blue square). The green square is the non-signaling region. The dark blue square represents the classical polytope for CHSH inequality in the ideal case (detectors always click). The red disk is the quantum polytope \cite{Branciard2011}. {\bf\ref{fig1b}} The shape of the polytope for detection efficiency $\eta=0.95$. There are parts of the quantum region (red) outside the classical one (blue). {\bf\ref{fig1c}} The shape of the polytope for detection efficiency $\eta=0.66$. The quantum region (red) is contained the classical one (blue). The symmetry of the classical polytope is broken because of the correlated assignments of Alice and Bob(both assign the same value). The symmetry would be broken in the other way if they assigned opposite values. {\bf\ref{fig1d}} The discard strategy. The region of the violation $\bar{I}_{3322}$ and $\bar{I}_{3322}^{sym}$ inequality for maximally entangled state as a function of the average detectors' efficiency. {\bf\ref{fig1e}} The assignment strategy. The region of the violation of $\bar{I}_{3322}$ inequality for a non-maximally entangled state as a function of the efficiency of detectors. The comparison of the strategies S00, S11 and S10.}
\end{figure*}
We have presented the influence of the choice of the post-selection strategy on CHSH inequality, which has a simple structure. However, more sophisticated Bell inequalities are more sensitive to the choice of post-selection strategy and entail more interesting effects.
We focus on $I_{3322}$ inequality \cite{Collins2004,Brunner2007,Horodecki2014} due do its asymmetry in the structure and consider the case in which the detection efficiencies are also asymmetric (different for Alice and Bob).

$I_{3322}$  inequality has different versions, which (in)equivalence again depends on the choice of the postprocessing strategy. The standard one reads:
\begin{eqnarray}
\bar{I}_{3322}&=&2P^{B}(1|0)+P^{B}(1|1)+P^{A}(1|0)+\\
&+&P(01|21)+P(10|21)+P(11|21)+\nonumber\\
&+&P(01|12)+P(10|12)+P(11|12)+\nonumber\\
&+&P(00|00)+P(00|10)+P(00|20)+\nonumber\\
&+&P(00|01)+P(00|02)+P(00|11)\ .\nonumber\label{e12}
\end{eqnarray}
For local realistic theories $\bar{I}_{3322}\leq 6$.

\subsection{Discard strategy for $\bar{I}_{3322}$ inequality}\label{ss41}
If one wants to examine the discard strategy for Bell inequality then one has to answer the following questions: Which representation of the inequality one has to consider? and Which detector strategy is optimal?

By the representation of the inequality one means its form depending on the way we calculate the marginal probabilities: $P^{B}(1|0)$, $P^{B}(1|1)$ and $P^{A}(1|0)$. Because the discard strategy can be signaling, each of these marginal probabilities can take a different value, for instance we can calculate $P^{B}(1|0)$ using one of the following:
\begin{eqnarray}
P^{B}(1|0)&=&P(01|00)+P(11|00)\label{e131}\\
&\text{or}&\nonumber\\
P^{B}(1|0)&=&P(01|10)+P(11|10)\label{e132}\\
&\text{or}&\nonumber\\
P^{B}(1|0)&=&P(01|20)+P(11|20)\label{e133}\ .
\end{eqnarray}
Hence, altogether there are nine representations of $\bar{I}_{3322}$ inequality.

By the detector strategy, one understands the probability of detector clicking for each of the settings in the optimal local model. $\bar{I}_{3322}$ inequality tests boxes with three inputs and two outputs, so the detector strategy for each party can be expressed as a three dimensional vector $(\eta_{1},\eta_{2},\eta_{3})$. Therefore, on average efficiency is equal $\eta_{T}=\frac{\eta_{1}+\eta_{2}+\eta_{3}}{3}$. It is easy to show that, the optimal classical strategy looks as follows: the detector clicks for two outputs with probability $1$ and for the remaining output with the probability $\eta$. Thus we only need to consider the average probability of clicking of the detectors in the range $\langle\frac{2}{3},1\rangle$.

Further calculations show that the lowest threshold efficiency is given by all representations of $\bar{I}_{3322}$ inequality in which marginal probabilities $P^{B}(1|0)$, $P^{B}(1|1)$ are calculated for Alice's setting $x$ different from $2$, i.e.:
\begin{eqnarray}
P^{B}(1|0)&=&P(01|x0)+P(11|x0)\label{e141}\ ,\\
P^{B}(1|1)&=&P(01|x0)+P(11|x0)\label{e142}\ ,
\end{eqnarray}
with $x\neq 2$.

The results obtained for $\bar{I}_{3322}$ were compared (fig. \ref{fig1d}) with ones obtained for the symmetric version of $\bar{I}_{3322}^{sym}$ \citep{Brunner2007a} (which, for assignment strategy is equivalent):

\begin{eqnarray}
\bar{I}_{3322}^{sym}&=&P^{A}(1|0)+P^{A}(1|1)+\\
&+&P^{B}(1|0)+P^{B}(1|1)+\nonumber\\
&+&P(01|11)+P(10|11)+P(11|11)+\nonumber\\
&+&P(01|22)+P(10|22)+P(21|22)+\nonumber\\
&+&P(00|10)+P(00|20)+P(00|01)+\nonumber\\
&+&P(00|02)+P(00|12)+P(00|21)\ .\nonumber
\end{eqnarray}

\subsection{Assignment strategy for $\bar{I}_{3322}$ inequality}\label{ss42}

In \citep{Brunner2007} authors showed that the assignment strategy applied to $I_{3322}$ inequality can reach the threshold detector's efficiency close to $43\%$. The main idea was to test pure non-maximally entangled states instead of maximally entangled. However, they examined the case in which one of two detectors is ideal according to the model of the entangled atom-photon pair. In the presented research both detectors are inefficient, which is a natural expansion of the mentioned model.

For the assignment strategy one determines threshold detector efficiency by solving the following inequality:
\begin{eqnarray}
\eta_{A}\eta_{B}Q_{\bar{I}_{3322}}&+&(1-\eta_{A})\eta_{B}W_{\bar{I}_{3322}}^{A}+\label{e3}\\+\eta_{A}(1-\eta_{B})W_{\bar{I}_{3322}}^{B}&+&(1-\eta_{A})(1-\eta_{B})C_{\bar{I}_{3322}}\nonumber\\
&\geq& C_{\bar{I}_{3322}}\nonumber \ .
\end{eqnarray}
In the above formula $Q_{\bar{I}_{3322}}$ is the quantum value of $\bar{I}_{3322}$ inequality, $W_{\bar{I}_{3322}}^{A}(W_{\bar{I}_{3322}}^{B})$
is the value of $\bar{I}_{3322}$ inequality when Alice's (Bob's) detector does not click and Bob's (Alice's) detector clicks, and $C_{\bar{I}_{3322}}$ is the maximal classical value of $\bar{I}_{3322}$ inequality. Analysing the formula \ref{e3} one can notice that it is impossible to reach value greater than $Q_{\bar{I}_{3322}}$. 

The optimal states and measurements for $\bar{I}_{3322}$ are not known, so to make our studies feasible in this paper we restrict ourselves to quantum strategies involving the states of the form $|\psi(\theta)\rangle=\cos{\theta}|00\rangle+\sin{\theta}|11\rangle$ and projective measurements. For these states a numerical optimisation was made. Similarly to \citep{Brunner2007} the optimal value was obtained for weakly entangled ones when $\theta=\frac{\Pi}{100}$.

Due to its symmetries, the behaviour of CHSH inequality does not depend on the choice of the outcomes assigned to the no clicking events. For  $\bar{I}_{3322}$ it is no longer the case. We have considered four possible types of assignment strategy ($\text{S}xy$, where $x,y\in\{0,1\}$ are Alice and Bob's assignments, respectively).

We find that the optimality of the type of strategy depends on the detection efficiencies of the parties. In most of the cases S10 is optimal, however there are small regions for which S00 or S11 are better (see fig. \ref{fig1e}). The worst strategy is S01. The region corresponding to that strategy lies inside the regions corresponding to any of the others.  

\section{Summary}
The main aim of the paper was to show the influence of the choice of post-processing method on Bell inequalities.

We showed that CHSH inequality is equivalent to CH and E inequalities if the assignment strategy is chosen, while inequivalent for discard. A similar effect appears for  $I_{3322}$ and its variants. Next we focused on the effects not present in the CHSH due to its simplicity. For $\bar{I}_{3322}$ inequality we have observed that the optimality of the assignment strategy depends on the type of the assignment. We have also shown that, in the case of the discard strategy, choosing a particular form for calculating marginal probabilities leads to different, inequivalent forms of the inequality.

In both cases we have found the assignment strategy to be a better choice than the discard. We conjecture it to be the case for any other Bell inequality.

\begin{acknowledgments}
This work is supported by FNP programme First TEAM (Grant No. First TEAM/2016-1/5).
\end{acknowledgments}
\clearpage
\bibliography{bibtex_bell_assymmetry}
\end{document}